# Montage: a grid portal and software toolkit for science-grade astronomical image mosaicking

**Joseph C. Jacob*, Daniel S. Katz**
Jet Propulsion Laboratory, California Institute of Technology,
Pasadena, CA 91109-8099
E-mail: {Joseph.C.Jacob*, Daniel.S.Katz}@jpl.nasa.gov
*Corresponding author

**G. Bruce Berriman, John Good, Anastasia C. Laity**
Infrared Processing and Analysis Center, California Institute of Technology
Pasadena, CA 91125
E-mail: {gbb, jcg, laity}@ipac.caltech.edu

**Ewa Deelman, Carl Kesselman, Gurmeet Singh, Mei-Hui Su**
USC Information Sciences Institute,
Marina del Rey, CA 90292
E-mail: {deelman, carl, gurmeet, mei}@isi.edu

**Thomas A. Prince, Roy Williams**
California Institute of Technology
Pasadena, CA 91125
E-mail: prince@srl.caltech.edu, roy@cacr.caltech.edu

**Abstract:** Montage is a portable software toolkit for constructing custom, science-grade mosaics by composing multiple astronomical images. The mosaics constructed by Montage preserve the astrometry (position) and photometry (intensity) of the sources in the input images. The mosaic to be constructed is specified by the user in terms of a set of parameters, including dataset and wavelength to be used, location and size on the sky, coordinate system and projection, and spatial sampling rate. Many astronomical datasets are massive, and are stored in distributed archives that are, in most cases, remote with respect to the available computational resources. Montage can be run on both single- and multi-processor computers, including clusters and grids. Standard grid tools are used to run Montage in the case where the data or computers used to construct a mosaic are located remotely on the Internet. This paper describes the architecture, algorithms, and usage of Montage as both a software toolkit and as a grid portal. Timing results are provided to show how Montage performance scales with number of processors on a cluster computer. In addition, we compare the performance of two methods of running Montage in parallel on a grid.

**Biographical notes:** Joseph C. Jacob has been a researcher at the Jet Propulsion Laboratory, California Institute of Technology, since he completed his PhD in computer engineering at Cornell University in 1996. His research interests are in the areas of parallel and distributed computing, image processing and scientific visualization.

Daniel S. Katz received his PhD in Electrical Engineering from Northwestern University in 1994. He is currently a Principal member of the Information Systems and Computer Science Staff at JPL. His current research interest include numerical methods, algorithms, and programming applied to supercomputing, parallel computing, and cluster computing; and fault-tolerant computing.

G. Bruce Berriman received his PhD in Astronomy from the California Institute of Technology. He is currently the project manager of NASA's Infrared Science Archive, based at the Infrared Processing and Analysis Center at the California Institute of Technology. His current research interests include searching for the very coolest stars ("brown dwarfs") and the development of on-request Grid services for astronomy.

John C. Good received his PhD in Astronomy from the University of Massachusetts in 1983. He is currently the project architect for NASA's Infrared Science Archive, based at the Infrared Processing and Analysis Center at the California Institute of Technology and has been heavily involved for many years in distributed information systems development for astronomy.

Anastasia Clower Laity received a Bachelor's Degree in Physics and Astronomy from Pomona College in 2000. She has been at the Infrared Processing and Analysis Center (IPAC) since 2002, working primarily as a software developer and system administrator on several data archiving projects.

Ewa Deelman is a Research Team Leader at the Center for Grid Technologies at the USC Information Sciences Institute and an Assistant Research Professor at the USC Computer Science Department. Dr. Deelman's research interests include the design and exploration of collaborative scientific environments based on Grid technologies, with particular emphasis on workflow management as well as the management of large amounts of data and metadata. Dr. Deelman received her PhD from Rensselaer Polytechnic Institute in Computer Science in 1997 in the area of parallel discrete event simulation.

Carl Kesselman is the Director of the Center for Grid Technologies at the Information Sciences Institute at the University of Southern California. He is an ISI Fellow and a Research Professor of Computer Science at the University of Southern California. Dr. Kesselman received his PhD in Computer Science at the University of California at Los Angeles and an honorary PhD from the University of Amsterdam.

Gurmeet Singh received his MS in Computer Science from University of Southern California in 2003. He is currently a graduate student in the Computer Science department at USC. His current research interests include resource provisioning and task scheduling aspects of grid computing.

Mei-Hui Su received her BS in Electrical Engineering and Computer Sciences from University of California, Berkeley. She currently is a system programmer at the Center for Grid Technologies at the USC Information Sciences Institute. Her interests include Grid technology, distributed and parallel computing.

Thomas A. Prince is a Professor of Physics at the California Institute of Technology. He received his PhD from the University of Chicago. Currently he is serving as the Chief Scientist of the NASA Jet Propulsion Laboratory. His research includes gravitational wave astronomy and computational astronomy.

Roy Williams received his PhD in Physics from the California Institute of Technology. He is currently a senior scientist at the Caltech Center for Advanced Computing Research. His research interests include science gateways for the grid, multi-plane analysis of astronomical surveys, and real-time astronomy.

## 1 INTRODUCTION

Wide-area imaging surveys have assumed fundamental importance in astronomy. They are being used to address such fundamental issues as the structure and organization of galaxies in space and the dynamical history of our galaxy. One of the most powerful probes of the structure and evolution of astrophysical sources is their behavior with wavelength, but this power has yet to be fully realized in the analysis of astrophysical images because survey results are published in widely varying coordinates, map projections, sizes and spatial resolutions. Moreover, the spatial extent of many astrophysical sources is much greater than that of individual images. Astronomy therefore needs a general image mosaic engine that will deliver image mosaics of arbitrary size in any common coordinate system, in any map projection and at any spatial sampling rate. The Montage project (Berriman et al., 2002; 2004) has developed this capability as a scalable, portable toolkit that can be used by astronomers on their desktops for science analysis, integrated into project and mission pipelines, or run on computing grids to support large-scale product generation, mission planning and quality assurance. Montage produces science-grade mosaics that preserve the photometric (intensity) and astrometric (location) fidelity of the sources in the input images.

Sky survey data are stored in distributed archives that are often remote with respect to the available computational resources. Therefore, state-of-the-art computational grid technologies are a key element of the Montage portal architecture. The Montage project is deploying a science-grade custom mosaic service on the Distributed Terascale Facility or TeraGrid (http://www.teragrid.org/). TeraGrid is a distributed infrastructure, sponsored by the National Science Foundation (NSF), and is capable of 20 teraflops peak performance, with 1 petabyte of data storage, and 40 gigabits per second of network connectivity between the multiple sites.

The National Virtual Observatory (NVO, http://www.us-vo.org/) and International Virtual Observatory Alliance (http://www.ivoa.net/) aim to establish the infrastructure necessary to locate, retrieve, and analyze astronomical data hosted in archives around the world. Science application portals can easily take advantage of this infrastructure by complying with the protocols for data search and retrieval that are being proposed and standardized by these virtual observatory projects. Montage is an example of one such science application portal being developed for the NVO.

Astronomical images are almost universally stored in Flexible Image Transport System (FITS) format (http://fits.gsfc.nasa.gov/). The FITS format encapsulates the image data with a meta-data header containing keyword-value pairs that, at a minimum, describe the image dimensions and how the pixels map to the sky. The World Coordinate System (WCS) specifies image-to-sky coordinate transformations for a number of different coordinate systems and projections useful in astronomy (Greisen and Calabretta, 2002). Montage uses FITS for both the input and output data format and WCS for specifying coordinates and projections.

Montage is designed to be applicable to a wide range of astronomical image data, and has been carefully tested on images captured by three prominent sky surveys spanning multiple wavelengths, the Two Micron All Sky Survey, 2MASS (http://www.ipac.caltech.edu/2mass/), the Digitized Palomar Observatory Sky Survey, DPOSS (http://www.astro.caltech.edu/~george/dposs/), and the Sloan Digital Sky Survey (SDSS). 2MASS has roughly 10 terabytes of images and catalogues (tabulated data that quantifies key attributes of selected celestial objects found in the images), covering nearly the entire sky at 1-arc-second sampling in three near-infrared wavelengths. DPOSS has roughly 3 terabytes of images, covering nearly the entire northern sky in one near-infrared wavelength and two visible wavelengths. The SDSS fourth data release (DR4) contains roughly 7.4 terabytes of images and catalogues covering 6,670 square degrees of the Northern sky in five visible wavelengths.

Two previously published papers provide background on Montage. The first described Montage as part of the architecture of the National Virtual Observatory (Berriman et al., 2002), and the second described some of the initial grid results of Montage (Berriman et al., 2004). In addition, a book chapter and a paper (Katz et al., 2005a; 2005b) provide highlights of the results reported in this paper. We extend these previous publications by providing additional details about the Montage algorithms, architectures, and usage.

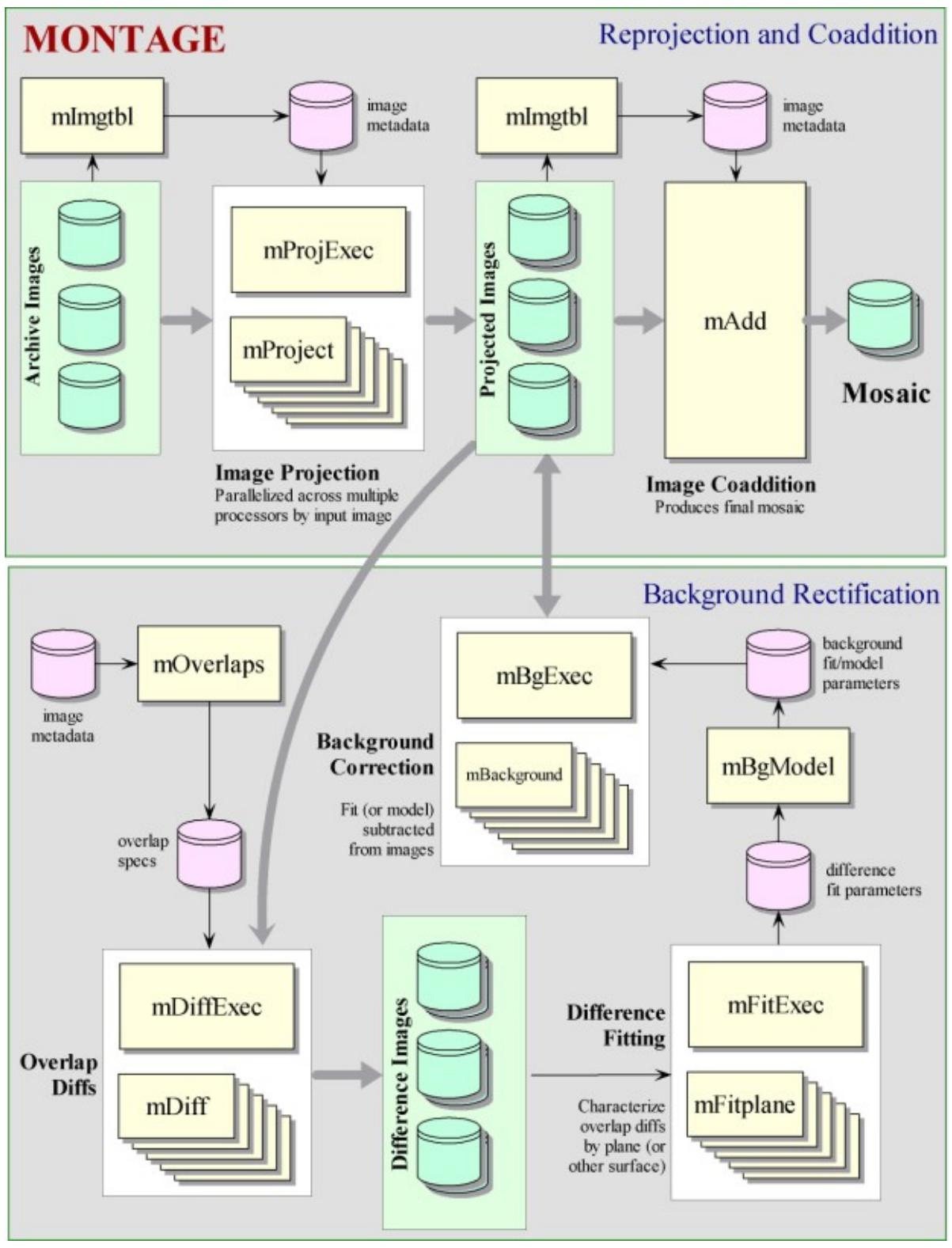

**Figure 1** *The high-level design of Montage.*

In this paper, we describe the modular Montage toolkit, the algorithms employed, and two strategies that have been used to implement an operational service on the TeraGrid, accessible through a web portal. The remainder of the paper is organized as follows. Section 2 describes how Montage is designed as a modular toolkit. Section 3 describes the algorithms employed in Montage. Section 4 describes the architecture of the Montage TeraGrid portal. Two strategies for running Montage on the TeraGrid are described in Sections 5 and 6, with a performance comparison in Section 7. A summary is provided in Section 8.

## 2 MONTAGE COMPONENTS

Montage's goal is to provide astronomers with software for the computation of custom science-grade image mosaics in FITS format. Custom refers to user specification of mosaic parameters, including WCS projection, coordinate system, mosaic size, image rotation, and spatial sampling rate. Science-grade mosaics preserve the calibration and astrometric (spatial) fidelity of the input images.

There are three steps to building a mosaic with Montage:

- Reprojection of input images to a common projection, coordinate system, and spatial scale,
- Modeling of background radiation in images to rectify them to a common flux scale and background level, thereby minimizing the inter-image differences, and
- Coaddition of reprojected, background-rectified images into a final mosaic.

Montage accomplishes these tasks in independent, portable, ANSI C modules. This approach controls testing and maintenance costs, and provides flexibility to users. They can, for example, use Montage simply to reproject sets of images and co-register them on the sky, implement a custom background removal algorithm, or define another processing flow through custom scripts. Table 1 describes the core computational Montage modules and Figure 1 illustrates how they may be used to produce a mosaic.

Three usage scenarios for Montage are as follows: the modules listed in Table 1 may be run as stand-alone programs; the executive programs listed in the table (i.e., *mProjExec*, *mDiffExec*, *mFitExec*, and *mBgExec*) may be used to process multiple input images either sequentially or in parallel via MPI; or the grid portal described in Section 4 may be used to process a mosaic in parallel on a computational grid. The modular design of Montage permits the same set of core compute modules to be used regardless of the computational environment being used.

## 3 MONTAGE ALGORITHMS

**Table 1** *The core design components of Montage*

| Component | Description |
|---|---|
| Mosaic Engine Components | |
| *mImgtbl* | Extract geometry information from a set of FITS headers and create a metadata table from it. |
| *mProject* | Reproject a FITS image. |
| *mProjExec* | A simple executive that runs *mProject* for each image in an image metadata table. |
| *mAdd* | Coadd the reprojected images to produce an output mosaic. |
| Background Rectification Components | |
| *mOverlaps* | Analyze an image metadata table to determine which images overlap on the sky. |
| *mDiff* | Perform a simple image difference between a pair of overlapping images. This is meant for use on reprojected images where the pixels already line up exactly. |
| *mDiffExec* | Run *mDiff* on all the overlap pairs identified by *mOverlaps*. |
| *mFitplane* | Fit a plane (excluding outlier pixels) to an image. Meant for use on the difference images generated by *mDiff*. |
| *mFitExec* | Run *mFitplane* on all the *mOverlaps* pairs. Creates a table of image-to-image difference parameters. |
| *mBgModel* | Modeling/fitting program which uses the image-to-image difference parameter table to interactively determine a set of corrections to apply to each image to achieve a "best" global fit. |
| *mBackground* | Remove a background from a single image (a planar correction has proven to be adequate for the images we have dealt with). |
| *mBgExec* | Run *mBackground* on all the images in the metadata table |

As described in Section 2, Montage constructs a mosaic through separate modules for image reprojection, background rectification, and coaddition. This section describes the main algorithms used in these modules.

### 3.1 General image reprojection

The first step in mosaic construction is to reproject each input image to the spatial scale, coordinate system, and projection of the output mosaic. Image reprojection involves the redistribution of information from a set of input pixels to a set of output pixels. For astronomical data, the input pixels represent the total energy received from an area on the sky, and it is critical to preserve this information when redistributed into output pixels. Also, in astronomy it is important to preserve the positional (astrometric) accuracy of the energy distribution, so common techniques such as adding all the energy from an input pixel to the "nearest" output pixel are inadequate. Instead, we must redistribute input pixel energy to the output based on the exact overlap of these pixels, possibly even using a weighting function across the pixels based on the point spread function for the original instrument.

The most common approach to determining pixel overlap is to project the input pixel into the output Cartesian space. This works well for some projection transformations but is difficult for others. One example of a difficult transformation is the Aitoff projection, commonly used in astronomy, where locations at the edge of an image correspond to undefined locations in pixel space. For Montage, we have decided instead to project both input and output pixels onto the celestial sphere. Since all such "forward" projections are well defined, the rest of the problem reduces to calculating the area of overlap of two convex polygons on a sphere (with no further consideration of the projections). The issue of handling reprojections thus becomes a problem of classical spherical trigonometry.

General algorithms exist for determining the overlap of polygons in Cartesian space (O'Rourke, 1998). We have modified this approach for use in spherical coordinates to determine the intersection polygon on the sphere (a convex hull) and applied Girard's Theorem, which gives the area of a spherical triangle based on the interior angles, to calculate the polygon's area.

The result is that for any two overlapping pixels, we can determine the area of the sky from the input pixel that contributes energy to the output pixel. This provides a mechanism for accurately distributing input energy to output pixels and a natural weighting mechanism when combining overlapping images.

Our approach implicitly assumes that the polygon defining a single pixel can be approximated by the set of great circle segments connecting the pixel's corners. Since even the largest pixels in any realistic image are on the order of a degree across, the nonlinearities along a pixel edge are insignificant. Furthermore, the only affect this would have would be to the astrometric accuracy of the energy location information and it would amount to a very small fraction (typically less that 0.01) of the size of a pixel. Total energy is still conserved.

### 3.2 Rapid image reprojection

Image reprojection is by far the most compute-intensive part of the processing because, in its general form, mapping from input image to output mosaic coordinates is done in two steps. First, input image coordinates are mapped to sky coordinates (i.e., right ascension and declination, analogous to longitude and latitude on the Earth). Second, those sky coordinates are mapped to output image coordinates. All of the mappings from one projection to another are compute-intensive, but some require more costly trigonometric operations than others and a few require even more costly iterative algorithms. The first public release of Montage applied this two-step process to the corners of the input pixels in order to map the flux from input image space to output space. Because the time required for this process stood as a serious obstacle to using Montage for large-scale image mosaics of the sky, a novel algorithm that is about 30 times faster was devised for the second code release.

The new much faster algorithm uses a set of linear equations (though not a linear transform) to transform directly from input pixel coordinates to output pixel coordinates. This alternate approach is limited to cases where both the input and output projections are "tangent plane" type (gnomonic, orthographic, etc.), but since these projections are by far the most common, it is appropriate to treat them as a special case.

This "plane-to-plane" approach is based on a library developed at the Spitzer Science Center (Makovoz and Khan, 2004). When both images are tangent plane, the geometry of the system can be viewed as in Figure 2, where a pair of gnomonic projection planes intersects the coordinate sphere. A single line connects the center of the sphere, the projected point on the first plane and the projected point on the second plane. This geometric relationship results in transformation equations between the two planar coordinate systems that require no trigonometry or extended polynomial terms. As a consequence, the transform is a factor of thirty or more faster than using the normal spherical projection formulae.

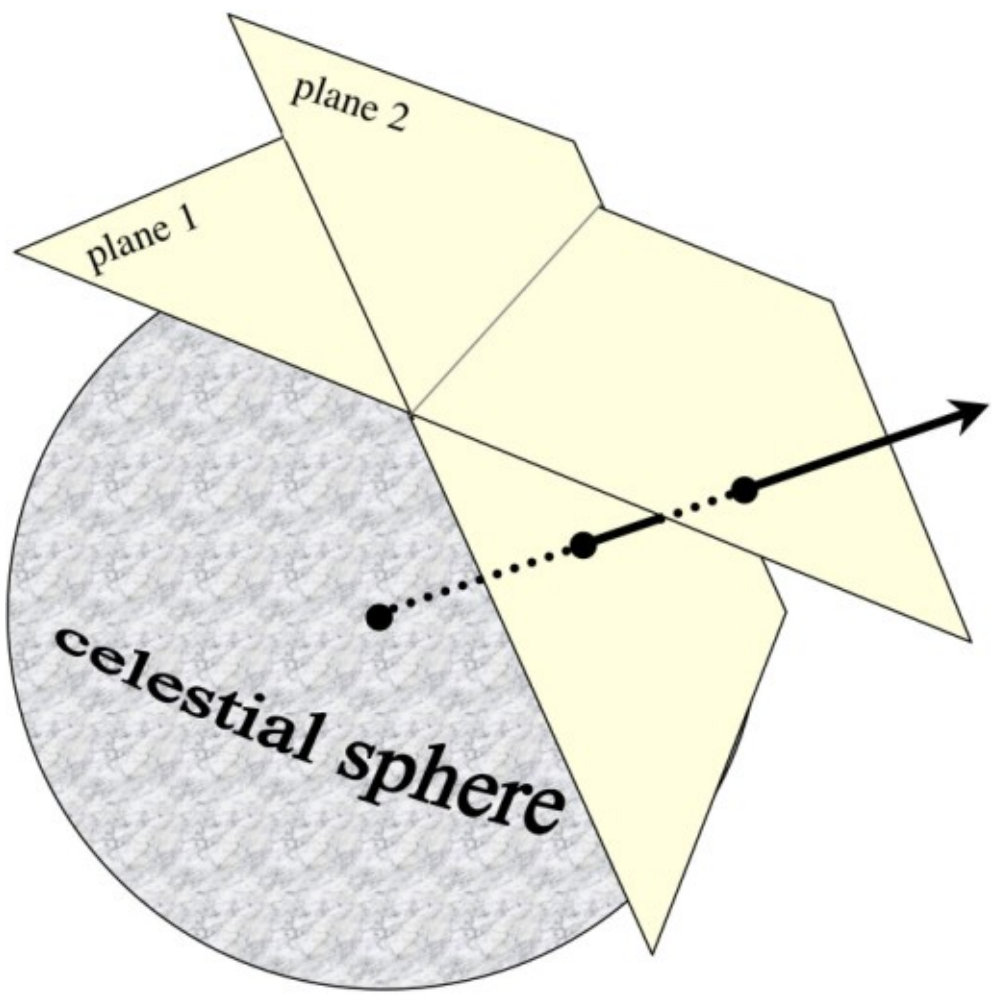

**Figure 2** *Plane-to-plane reprojection.*

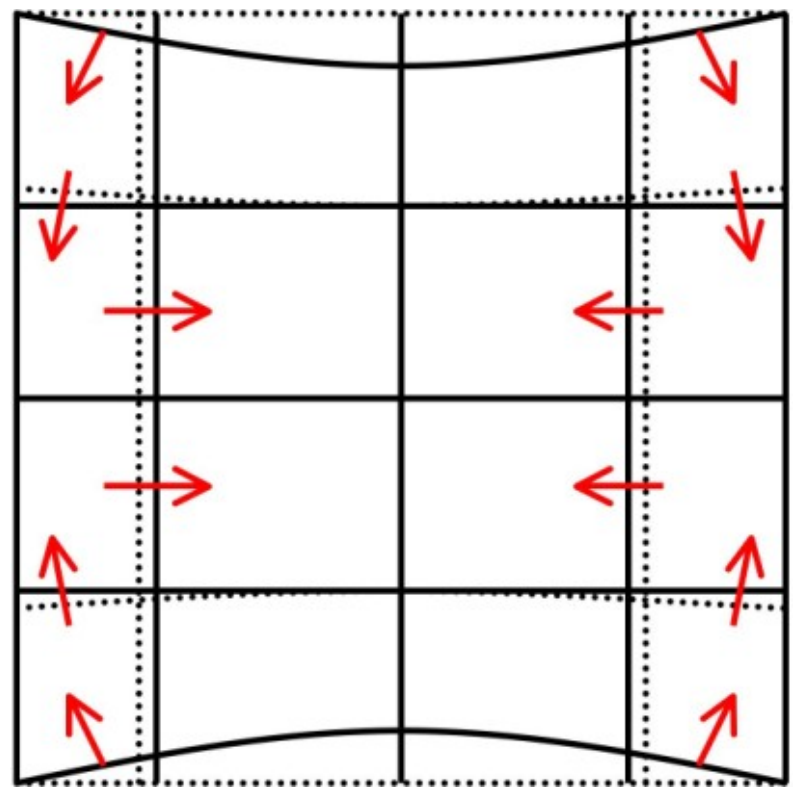

**Figure 3** *Representation of a WCS projection as a distorted Gnomonic (TAN) projection, exaggerated for clarity. The arrows indicate the sense of the distortions.*

A bonus to the plane-to-plane approach is that the computation of pixel overlap is much easier, involving only clipping constraints of the projected input pixel polygon in the output pixel space.

This approach excludes many commonly-used projections such as "Cartesian" and "zenithal equidistant," and is essentially limited to small areas of few square degrees. Processing of all-sky images, as is almost always the case with projections such as Aitoff, generally requires the slower plane-to-sky-to-plane approach.

There is, however, a technique that can be used for images of high resolution and relatively small extent (up to a few degrees on the sky). Rather than use the given image projection, we can often approximate it with a very high degree of accuracy with a "distorted" Gnomonic projection. In this case, the pixel locations are "distorted" by small distances relative to the plane used in the image projection formulae. A distorted space is one in which the pixel locations are slightly offset from the locations on the plane used by the projection formulae, as happens when detectors are slightly misshapen, for instance. This distortion is modelled by pixel-space polynomial correction terms that are stored as parameters in the image FITS header.

While this approach was developed to deal with physical distortions caused by telescope and instrumental effects, it is also applicable to Montage in augmenting the plane-to-plane reprojection. Over a small, well-behaved region, most projections can be approximated by a Gnomonic (TAN) projection with small distortions. For instance, in terms of how pixel coordinates map to sky coordinates, a two-degree "Cartesian" (CAR) projection is identical to a TAN projection with a fourth-order distortion term to within about a percent of a pixel width. Figure 3 shows this, in exaggerated form for clarity, with the arrows showing the sense of the distortion.

### 3.3 Background rectification

If several images are to be combined into a mosaic, they must all be projected onto a common coordinate system (see above) and then any discrepancies in brightness or background must be removed, as illustrated in Figure 4. Our assumption is that the input images are all calibrated to an absolute energy scale (i.e., brightnesses are absolute and should not be modified) and that any discrepancies between the images are due to variations in their background levels that are terrestrial or instrumental in origin.

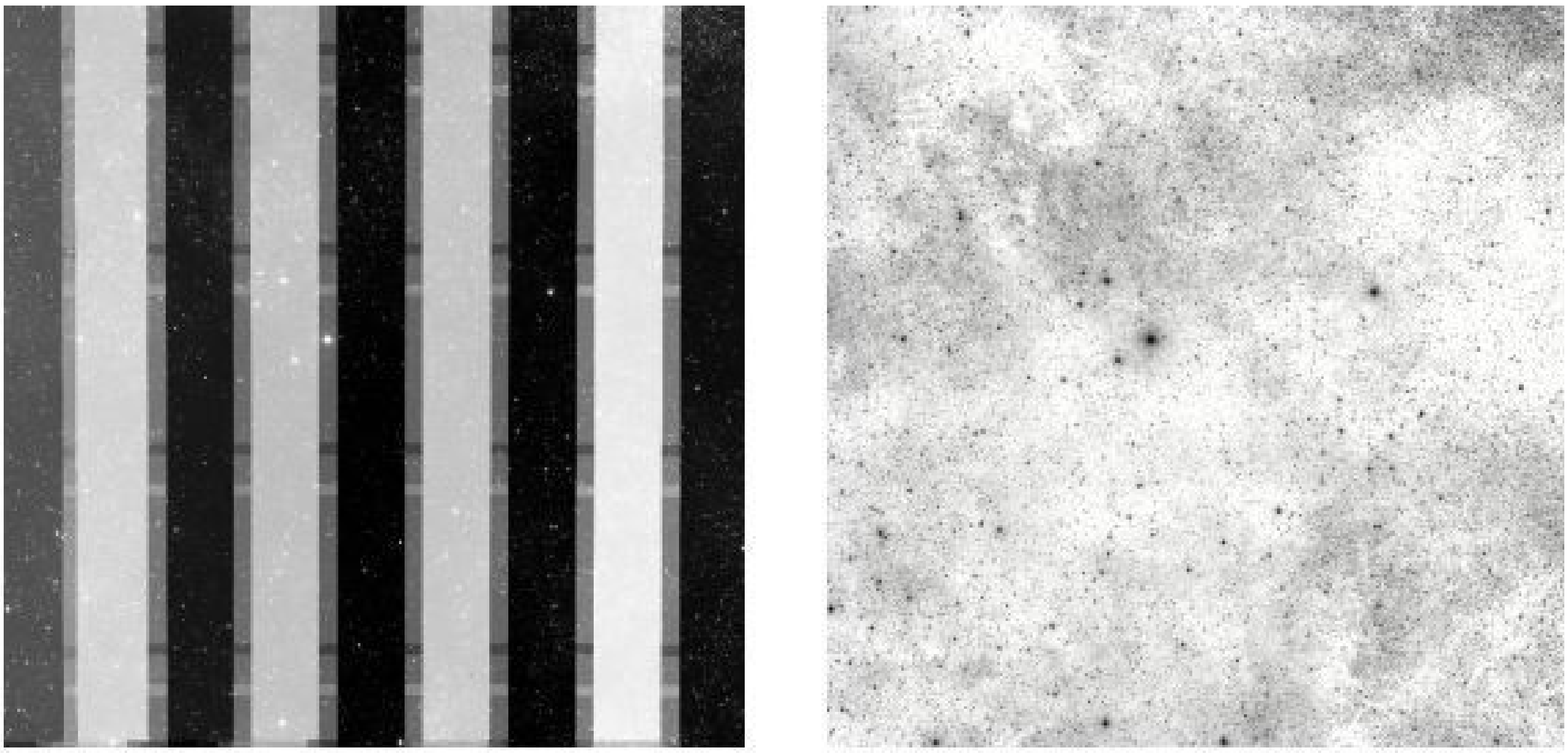

**Figure 4** *A Montage mosaic before (left) and after (right) background rectification.*

The Montage background rectification algorithm is based on the assumption that terrestrial and instrumental backgrounds can be described by simple linear functions or surfaces (e.g., slopes and offsets). Stated more generally, we assume that the "non-sky" background has very little energy in any but the lowest spatial frequencies. If this not the case, it is unlikely that any generalized background matching algorithm will be able distinguish between "sky" and rapidly varying "background"; background removal would then require an approach that depends on a detailed knowledge of an individual data set.

Given a set of overlapping images, characterization of the overlap differences is key to determining how each image should be adjusted before combining them. We consider each image individually with respect to its neighbours. Specifically, we determine the areas of overlap between each image and its neighbours, and use the complete set of overlap pixels in a least-squares fit to determine how each image should be adjusted (e.g., what gradient and offset should be added) to bring it best in line with its neighbours.

In practice, we adjust the image by half this optimal amount, since all the neighbors are also being analyzed and adjusted and we want to avoid ringing. After doing this for all the images, we iterate (currently for a fixed number of iterations, though a convergence criteria could be used). The final effect is to have subtracted a low-frequency (currently a gradient and offset) background from each image such that the cumulative image-to-image differences are minimized. To speed the computation and minimize memory usage, we approximate the gradient and offset values by a planar surface fit to the overlap area difference images rather than perform a least squares fit using all of the overlap pixels.

### 3.4 Coaddition

In the reprojection algorithm (described above), each input pixel's energy contribution to an output pixel is added to that pixel, weighted by the sky area of the overlap. In addition, a cumulative sum of these sky area contributions is kept for the output pixels (called an "area" image). When combining multiple overlapping images, these area images provide a natural weighting function; the output pixel value is simply the area-weighted average of the pixels being combined.

Such images are in practice very flat (with only slight slopes due to the image projection) since the cumulative effect is that each output pixel is covered by the same amount of input area, regardless of the pattern of coverage. The only real variation occurs at the edges of the area covered, since there an output pixel may be only partially covered by input pixels.

The limitations of available memory have been simply overcome in coaddition by reading the reprojected images one line at a time from files that reside on disk. Assuming that a single row of the output file does not fill the memory, the only limitation on file size is that imposed by the file system. Indeed, images of many gigabytes have thus far been built with the new software. For each output line, *mAdd* determines which input files will be contributing pixel values, and opens only those files. Each contributing pixel value is read from the flux and area coverage files, and the value of each of these pixels is stored in an array until all contributing pixels have been read for the corresponding output row. This array constitutes a "stack" of input pixel values; a corresponding stack of area coverage values is also preserved. The contents of the output row are then calculated one output pixel (i.e., one input stack) at a time, by averaging the flux values from the stack.

Different algorithms to perform this average can be trivially inserted at this point in the program. Montage currently supports mean and median coaddition, with or without weighting by area. The mean algorithm (the default) accumulates flux values contributing to each output pixel, and then scales them by the total area coverage for that pixel. The median algorithm ignores any pixels whose area coverage falls below a specific threshold, and then calculates the median flux value from the remainder.

If there are no area files, then the algorithm gives equal weight to all pixels. This is valuable for science data sets where the images are already projected into the same pixel space (e.g., MSX). An extension of the algorithm to support outlier rejection is planned for a future release.

### 3.5 Drizzle

The Space Telescope Science Institute (STScI) has developed a method for the linear reconstruction of an image from under-sampled, dithered data. The algorithm is known as "drizzling," or more formally as Variable-Pixel Linear Reconstruction (Fruchter and Hook, 2002). Montage provides drizzle as an option in the image reprojection. In this algorithm, pixels in the original input images are mapped onto the output mosaic as usual, except the pixel is first "shrunken" by a user-defined amount. This is particularly easy to do in Montage. Since the Montage algorithm projects the corners of each pixel onto the sky, we implement drizzle by simply using a different set of corners in the interior of the original pixel. In other words, the flux is modelled as all coming from a box centered on the original pixel but smaller by the drizzle factor.

## 4 MONTAGE GRID PORTAL ARCHITECTURE

The basic user interface to Montage is implemented as a web portal. In this portal, the user selects a number of input parameters for the mosaicking job, such as the centre and size of the region of interest, the source of the data to be mosaicked, and some identification such as an email address. Once this information is entered, the user assumes that the mosaic will be computed, and she will be notified of the completion via an email message containing a URL where the mosaic can be retrieved.

Behind the scenes, a number of things have to happen. First, a set of compute resources needs to be chosen. Here, we will assume that this is a cluster with processors that have access to a shared file system. Second, the input data files and executable code needs to be moved to these resources. Third, the modules need to be executed in the right order. In general, this might involve moving intermediate files from one set of resources to another, but the previous assumption makes this file movement unnecessary. Fourth, the output mosaic and some status information need to be moved to a location accessible to the user. Fifth and finally, the user must be notified of the job completion and the location of the output files.

The Montage TeraGrid portal includes components distributed across computers at the Jet Propulsion Laboratory (JPL), Infrared Processing and Analysis Center (IPAC), USC Information Sciences Institute (ISI), and the TeraGrid, as illustrated in Figure 5. Note that the description here applies to 2MASS mosaics, but can be easily extended to DPOSS and SDSS images as well. The portal is comprised of the following five main components, each having a client and server as described below: (i) User Portal, (ii) Abstract Workflow Service, (iii) 2MASS Image List Service, (iv) Grid Scheduling and Execution Service, and (v) User Notification Service.

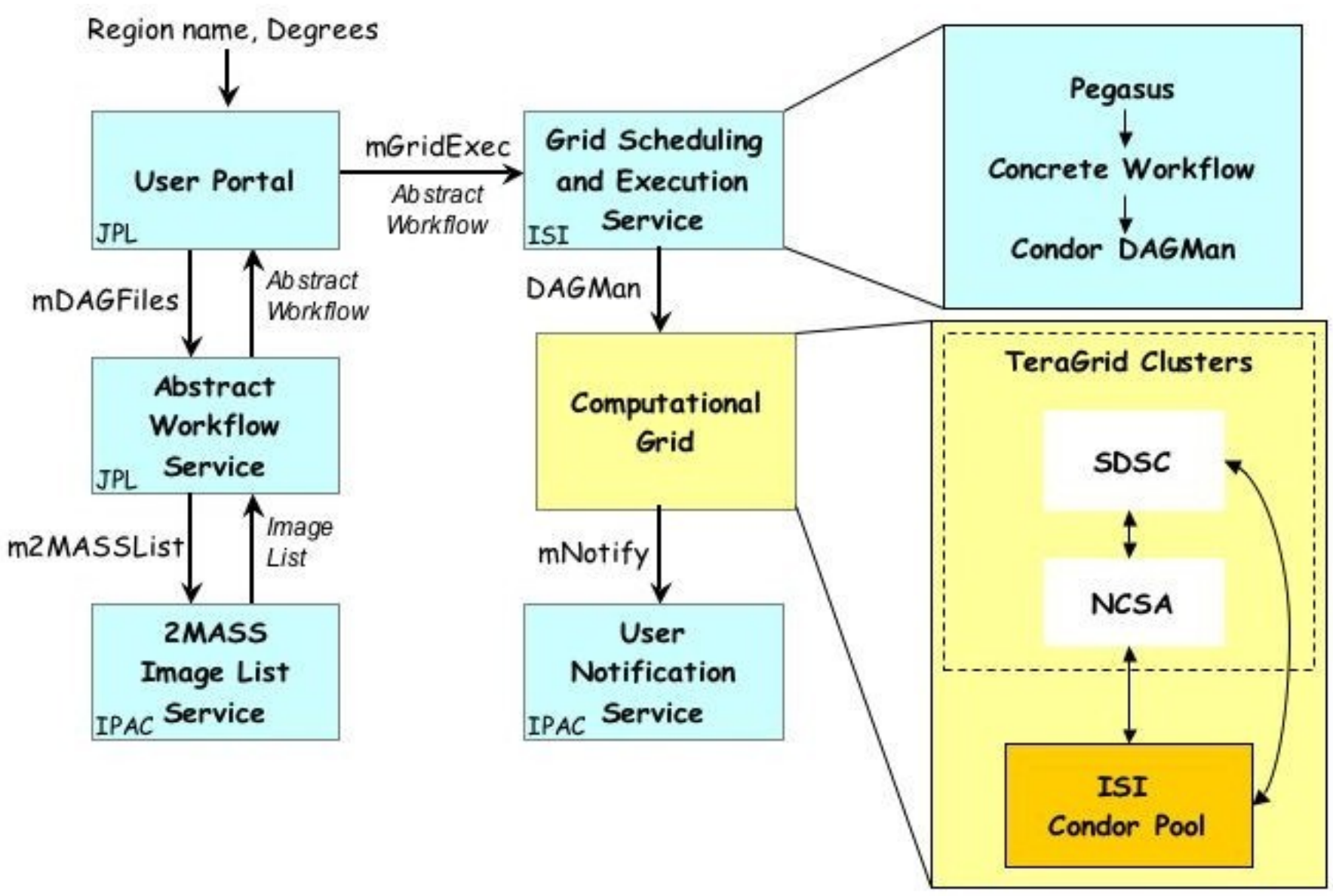

**Figure 5** *The Montage TeraGrid portal architecture.*

This design exploits the parallelization inherent in the Montage architecture. The Montage grid portal is flexible enough to run a mosaic job on a number of different cluster and grid computing environments, including Condor pools and TeraGrid clusters. We have demonstrated processing on both a single cluster configuration and on multiple clusters at different sites having no shared disk storage.

### 4.1 User portal

Users on the Internet submit mosaic requests by filling in a simple web form with parameters that describe the mosaic to be constructed, including an object name or location, mosaic size, coordinate system, projection, and spatial sampling rate. After request submission, the remainder of the data access and mosaic processing is fully automated with no user intervention.

The server side of the user portal includes a CGI program that receives the user input via the web server, checks that all values are valid, and stores the validated requests to disk for later processing. A separate daemon program with no direct connection to the web server runs continuously to process incoming mosaic requests. The processing for a request is done in two main steps:

1. Call the abstract workflow service client code
2. Call the grid scheduling and execution service client code and pass to it the output from the abstract workflow service client code

### 4.2 Abstract workflow service

The abstract workflow service takes as input a celestial object name or location on the sky and a mosaic size and returns a zip archive file containing the abstract workflow as a directed acyclic graph (DAG) in XML and a number of input files needed at various stages of the Montage mosaic processing. The abstract workflow specifies the jobs and files to be encountered during the mosaic processing, and the dependencies between the jobs. These dependencies are used to determine which jobs can be run in parallel on multiprocessor systems. A pictorial representation of an abstract workflow for computing a mosaic from three input images is shown in Figure 6.

### 4.3 2MASS image list service

The 2MASS Image List Service takes as input a celestial object name or location on the sky (which must be specified as a single argument string), and a mosaic size. The 2MASS images that intersect the specified location on the sky are returned in a table, with columns that include the filenames and other attributes associated with the images.

### 4.4 Grid scheduling and execution service

The Grid Scheduling and Execution Service takes as input the abstract workflow, and all of the input files needed to construct the mosaic. The service authenticates users using grid security credentials stored in a MyProxy server (Novotny et al., 2001), schedules the job on the grid using Pegasus (Deelman et al., 2002; 2003; 2004), and then executes the job using Condor DAGMan (Frey et al., 2001). Section 6 describes how this is done in more detail.

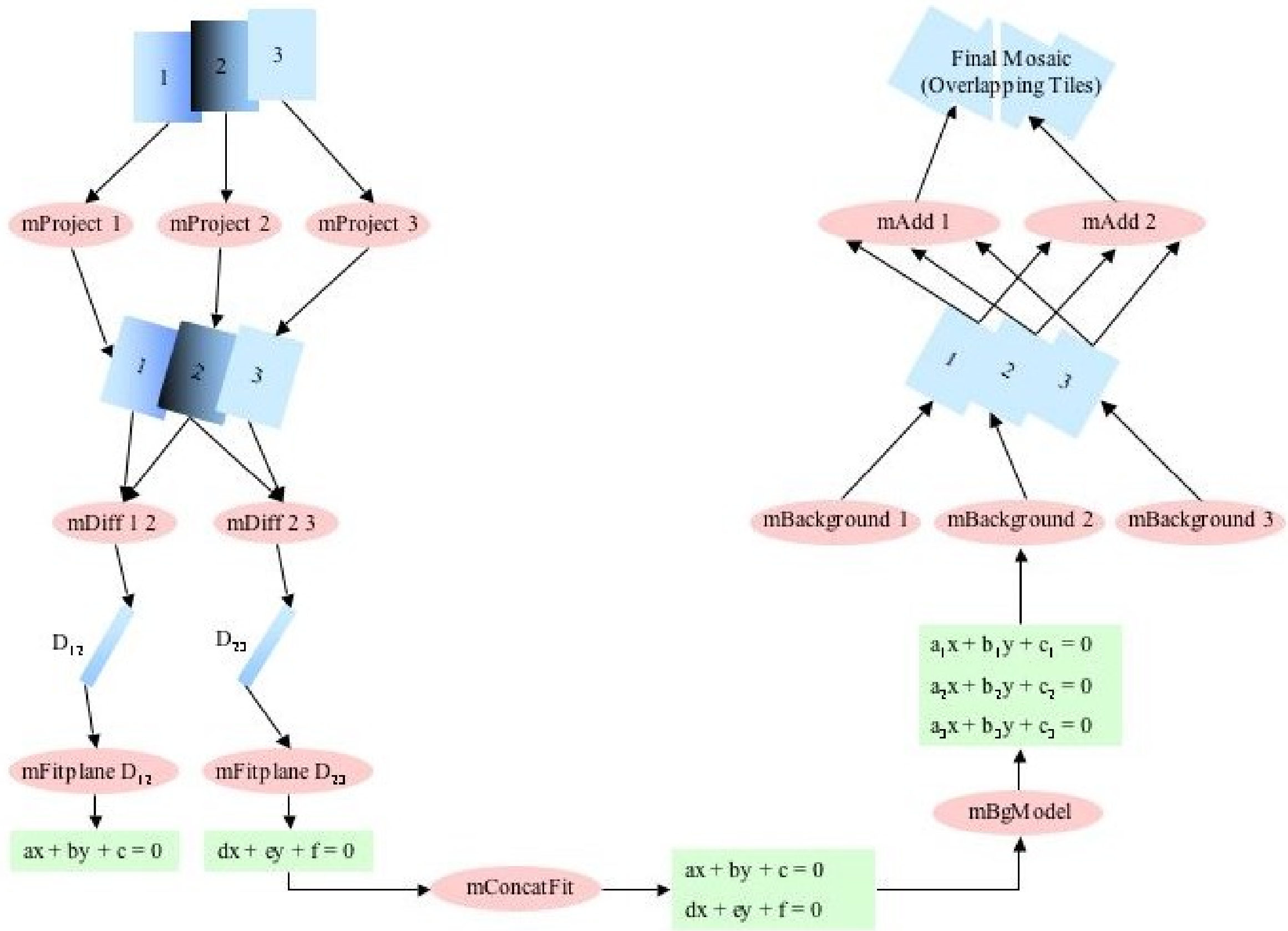

**Figure 6** *Example abstract workflow.*

### 4.5 User notification service

The last step in the grid processing is to notify the user with the URL where the mosaic may be downloaded. This notification is done by a remote user notification service at IPAC so that a new notification mechanism can be used later without having to modify the Grid Scheduling and Execution Service. Currently the user notification is done with a simple email, but a later version will use the Request Object Management Environment (ROME), being developed separately for the National Virtual Observatory. ROME will extend our portal with more sophisticated job monitoring, query, and notification capabilities.

## 5 GRID-ENABLING MONTAGE VIA MPI PARALLELIZATION

The first method of running Montage on a grid is to use grid-accessible clusters, such as the TeraGrid. This is very similar to traditional, non-grid parallelization. By use of MPI, the Message Passing Interface (Snir et al., 1996), the executives (*mProjExec*, *mDiffExec*, *mFitExec*, *mBgExec*, *mAddExec*) and *mAdd* can be run on multiple processors. The Atlasmaker (Williams et al., 2003) project previously developed an MPI version of *mProject*, but it was not closely coupled to the released Montage code, and therefore has not continued to work with current Montage releases. The current MPI versions of the Montage modules are generated from the same source code as the single-processor modules by preprocessing directives.

The structure of the executives are similar to each other, in that each has some initialization that involves determining a list of files on which a module will be run, a loop in which the module is actually called for each file, and some finalization work that includes reporting on the results of the module runs. The executives, therefore, are parallelized straightforwardly, with all processes of a given executive being identical to each other. All the initialization is duplicated by all of the processes. A line is added at the start of the main loop, so that each process only calls the sub-module if the remainder of the loop count divided by the number of processes equals the MPI rank (a logical identifier of an MPI process). All processes then participate in global sums to find the total statistics of how many sub-modules succeeded, failed, etc., as each process keeps track of its own statistics. After the global sums, only the process with rank 0 prints the global statistics.

*mAdd* writes to the output mosaic one line at a time, reading from its input files as needed. The sequential *mAdd* writes the FITS header information into the output file before starting the loop on output lines. In the parallel *mAdd*, only the process with rank 0 writes the FITS header information, then it closes the file. Each process then carefully seeks and writes to the correct part of the output file. Each process is assigned a unique subset of the rows of the mosaic to write, so there is no danger of one process overwriting the work of another. While the executives were written to divide the main loop operations in a round-robin fashion, it makes more sense to parallelize the main *mAdd* loop by blocks, since it is likely that a given row of the output file will depend on the same input files as the previous row, and this can reduce the amount of input I/O for a given process.

Note that there are two modules that can be used to build the final output mosaic, *mAdd* and *mAddExec*, and both can be parallelized as discussed in the previous two paragraphs. At this time, we have just run one or the other, but it would be possible to combine them in a single run.

A set of system tests is available from the Montage web site. These tests, which consist of shell scripts that call the various Montage modules in series, were designed for the single-processor version of Montage. The MPI version of Montage is run similarly, by changing the appropriate lines of the shell script, for example, from:

*mProjExec arg1 arg2 ...*

to:

*mpirun -np N mProjExecMPI arg1 arg2 ...*

No other changes are needed. If this is run on a queue system, a set of processors is reserved for the job. Some parts of the job, such as *mImgtbl*, only use one processor, and other parts, such as *mProjExecMPI*, use all the processors. Overall, most of the processors are in use most of the time. There is a small bit of overhead here in launching multiple MPI jobs on the same set of processors. One might change the shell script into a parallel program, perhaps written in C or Python, to avoid this overhead, but this has not been done for Montage.

The processing part of this approach is not very different from what might be done on a cluster that is not part of a grid. In fact, one might choose to run the MPI version of Montage on a local cluster by logging into the local cluster, transferring the input data to that machine, submitting a job that runs the shell script to the queuing mechanism, and finally, after the job has run, retrieving the output mosaic. Indeed, this is how the MPI code discussed in this paper was run and measured. The discussion of how this code could be used in a portal is believed to be correct, but has not been implemented and tested.

## 6 GRID-ENABLING MONTAGE WITH PEGASUS

Pegasus (Planning for Execution in Grids), developed as part of the GriPhyN Virtual Data (http://www.griphyn.org/),

is a framework that enables the mapping of complex workflows onto distributed resources such as the grid. In particular, Pegasus maps an "abstract workflow" to a "concrete workflow" that can be executed on the grid using a variety of computational platforms, including single hosts, Condor pools, compute clusters, and the TeraGrid.

An abstract workflow describes a computation in terms of logical transformations and data without identifying the resources needed to execute it. The Montage abstract workflow consists of the various application components shown in Figure 6. The nodes represent the logical transformations such as *mProject*, *mDiff* and others. The edges represent the data dependencies between the transformations. For example, *mConcatFit* requires all the files generated by all the previous *mFitplane* steps.

### 6.1 Mapping application workflows

Pegasus maps an abstract workflow description to a concrete, executable form after consulting various grid information services to find suitable resources, the data that is used in the workflow, and the necessary software. In addition to specifying computation on grid resources, this concrete, executable workflow also has data transfer nodes (for both stage-in and stage-out of data), data registration nodes that can update various catalogues on the grid (for example, RLS), and nodes that can stage-in binaries.

Pegasus finds any input data referenced in the workflow by querying the Globus Replica Location Service (RLS), assuming that data may be replicated across the grid (Chervenak et al., 2002). After Pegasus derives new data products, it can register them into the RLS as well.

Pegasus finds the programs needed to execute a workflow, including their environment setup requirements, by querying the Transformation Catalogue (TC) (Deelman et al., 2001). These executable programs may be distributed across several systems.

Pegasus queries the Globus Monitoring and Discovery Service (MDS) to find the available compute resources and their characteristics such as the load, the scheduler queue length, and available disk space (Czajkowski et al., 2001). Additionally, the MDS is used to find information about the location of the GridFTP servers (Allcock et al., 2002) that can perform data movement, job managers (Czajkowski et al., 2001) that can schedule jobs on the remote sites, storage locations, where data can be pre-staged, shared execution directories, the RLS into which new data can be registered, site-wide environment variables, etc.

The information from the TC is combined with the MDS information to make scheduling decisions, with the goal of scheduling a computation close to the data needed for it. One other optimization that Pegasus performs is to reuse those workflow data products that already exist and are registered into the RLS, thereby eliminating redundant computation. As a result, some components from the abstract workflow may not appear in the concrete workflow.

### 6.2 Workflow execution

The concrete workflow produced by Pegasus is in the form of submit files that are given to DAGMan and Condor-G for execution. The submit files indicate the operations to be performed on given remote systems and dependencies, to be enforced by DAGMAN, which dictate the order in which the operations need to be performed.

In case of job failure, DAGMan can retry a job a given number of times. If that fails, DAGMan generates a rescue workflow that can be potentially modified and resubmitted at a later time. Job retry is useful for applications that are sensitive to environment or infrastructure instability. The rescue workflow is useful in cases where the failure was due to lack of disk space that can be reclaimed or in cases where totally new resources need to be assigned for execution. Obviously, it is not always beneficial to map and execute an entire workflow at once, because resource availability may change over time. Therefore, Pegasus also has the capability to map and then execute (using DAGMan) one or more portions of a workflow (Deelman et al., 2004).

## 7 COMPARISON OF GRID EXECUTION STRATEGIES AND PERFORMANCE

Here we discuss the advantages and disadvantages of each of the two approaches (MPI and Pegasus) we took to running Montage on the grid. We quantify the performance of the two approaches and describe how they differ.

### 7.1 Benchmark problem and system

In order to test the two approaches to grid-enabling Montage, we chose a sample problem that could be computed on a single processor in a reasonable time as a benchmark. The results in this paper involve this benchmark, unless otherwise stated.

The benchmark problem generates a mosaic of 2MASS data from a 6 x 6 degree region at M16. This requires 1,254 input 2MASS images, each about 0.5 megapixel, for a total of about 657 megapixels (about 5 GB with 64 bits/pixel double precision floating point data). The output is a 3.7 GB FITS (Flexible Image Transport System) file with a 21,600 x 21,600 pixel data segment, and 64 bits/pixel double precision floating point data. The output data is a little smaller than the input data size because there is some overlap between neighboring input images. For the timing results reported in this section, the input data had been pre-staged to a local disk on the compute cluster.

Results in this paper are measured on the "Phase 2" TeraGrid cluster at the National Center for Supercomputing Applications (NCSA), unless otherwise mentioned. This cluster has (at the time of this experiment) 887 nodes, each with dual Itanium 2 processors with at least 4 GB of memory. 256 of the nodes have 1.3 GHz processors, and the other 631 nodes have 1.5 GHz processors. The timing tests reported in this paper used the 1.5 GHz processors. The network between nodes is Myrinet and the operating system is SuSE Linux. Disk I/O is to a 24 TB General Parallel File System (GPFS). Jobs are scheduled on the system using Portable Batch System (PBS) and the queue wait time is not included in the execution times since that is heavily dependent on machine load from other users.

Figure 6 shows the processing steps for the benchmark problem. There are two types of parallelism: simple file-based parallelism, and more complex module-based parallelism. Examples of file-based parallelism are the *mProject* modules, each of which runs independently on a single file. *mAddExec*, which is used to build an output mosaic in tiles, falls into this category as well, as once all the background-rectified files have been built, each output tile may be constructed independently, except for I/O contention. The second type of parallelism can be seen in *mAdd*, where any number of processors can work together to build a single output mosaic. This module has been parallelized over blocks of rows of the output, but the parallel processes need to be choreographed to write the single output file correctly. The results in this paper are for the serial version of *mAdd*, where each output tile is constructed by a single processor.

### 7.2 Starting the job

In both the MPI and Pegasus implementations, the user can choose from various sets of compute resources. For MPI, the user must specify a single set of processors that share a file system. For Pegasus, this restriction is removed since it can automatically transfer files between systems. Thus, Pegasus is clearly more general. Here, we compare performance on a single set of processors on the TeraGrid cluster, described previously as the benchmark system.

### 7.3 Data and code stage-in

In either approach, the need for both data and code stage-in is similar. The Pegasus approach has clear advantages, in that the RLS and Transformation Catalogue can be used to locate the input data and proper executables for a given machine, and can stage the code and data to an appropriate location. In the MPI approach, the user must know where the executable code is, which is not a problem when the portal executes the code, as it then is the job of the portal creator. Data reuse can also be accomplished with a local cache, though this is not as general as the use of RLS.

In any event, input data will sometimes need to be retrieved from an archive. In the initial version of the portal discussed in this paper, we use the 2MASS list service at IPAC, but a future implementation will use the proposed standard Simple Image Access (SIA) protocol (http://www.ivoa.net/Documents/latest/SIA.html), which returns a table listing the files (URLs) that can be retrieved.

### 7.4 Building the mosaic

With the MPI approach, the portal generates a shell script and a job to run it is submitted to the queue. Each command in the script is either a sequential or parallel command to run a step of the mosaic processing. The script will have some queue delay, and then will start executing. Once it starts, it runs until it finishes with no additional queue delays. The script does not contain any detail on the actual data files, just the directories. The sequential commands in the script examine the data directories and instruct the parallel jobs about the actual file names.

The Pegasus approach differs in that the initial work is more complex, but the work done on the compute nodes is much simpler. For reasons of efficiency, a pool of processors is allocated from the parallel machine by use of the queue. Once this pool is available, Condor-Glidein (http://www.cs.wisc.edu/condor/glidein/) is used to associate this pool with an existing Condor pool. Condor DAGMan then can fill the pool and keep it as full as possible until all the jobs have been run. The decision about what needs to be run and in what order is made by the portal, where the *mDAG* module builds the abstract DAG, and Pegasus then builds the concrete DAG.

Because the queuing delays are one-time delays for both methods, we do not discuss them any further. The elements for which we discuss timings below are the sequential and parallel jobs for the MPI approach, and the *mDAG*, Pegasus, and compute modules for the Pegasus approach.

### 7.5 MPI timing results

The timing results of the MPI version of Montage are shown in Figure 7. The total times shown in this figure include both the parallel modules (the times for which are also shown in the figure) and the sequential modules (the times for which are not shown in the figure, but are relatively small).

The end-to-end runs of Montage involved running the modules in this order: *mImgtbl, mProjExec, mImgtbl, mOverlaps, mDiffExec, mFitExec, mBgModel, mBgExec, mImgtbl, mAddExec*.

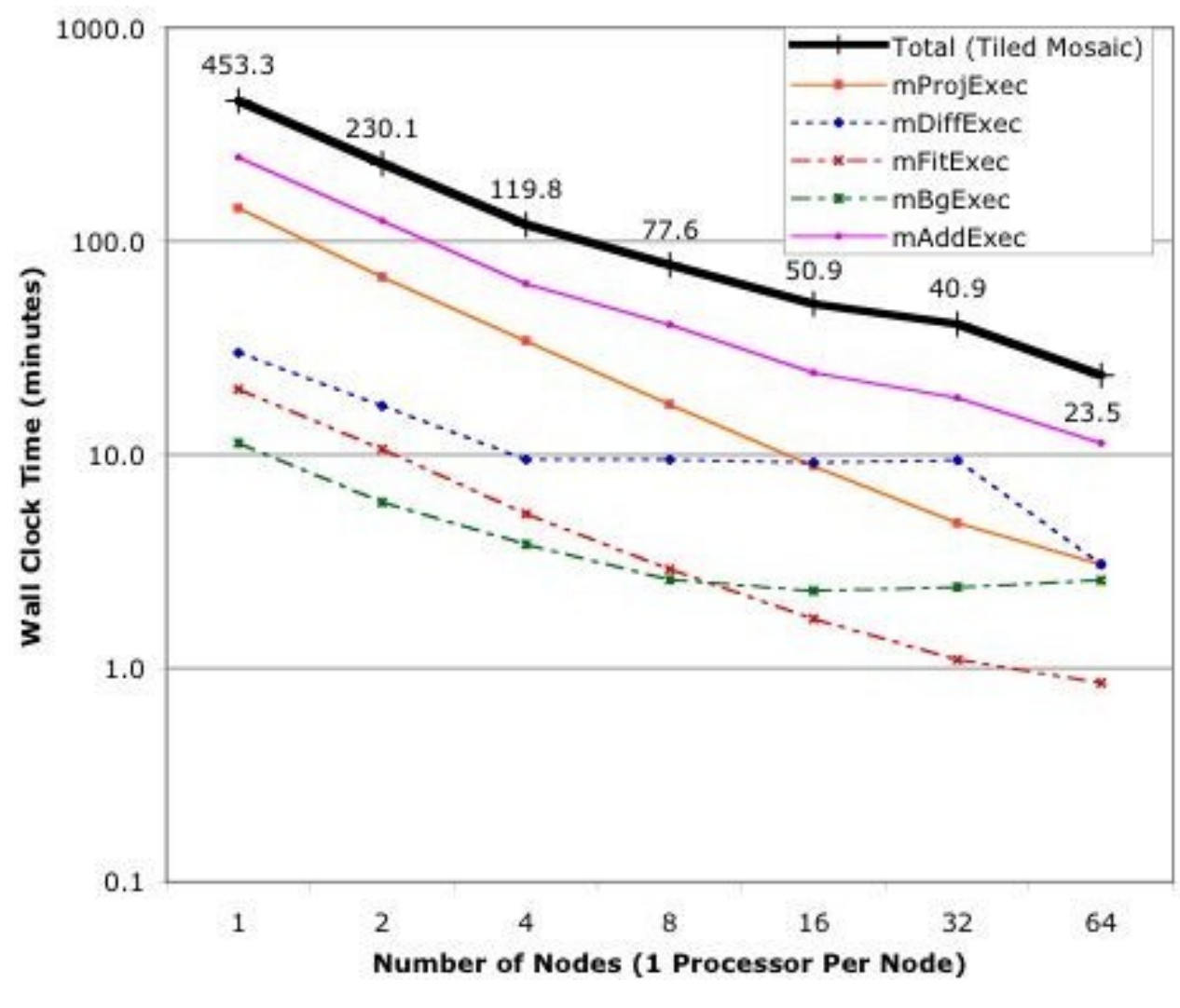

**Figure 7** *Performance of MPI version of Montage building a 6 x 6 degree mosaic.*

MPI parallelization reduces the one processor time of 453 minutes down to 23.5 minutes on 64 processors, for a speedup of 19. Note that with the exception of some small initialization and finalization code, all of the parallel code is non-sequential. The main reason the parallel modules fail to scale linearly as the number of processors is increased is I/O. On a system with better parallel I/O, one would expect to obtain better speedups; the situation where the amount of work is too small for the number of processors has not been reached, nor has the Amdahl's law limit been reached.

Note that there is certainly some variability inherent in these timings, due to the activity of other users on the cluster. For example, the time to run a serial module like *mImgtbl* shouldn't vary with number of processors, but the measured results vary from 0.7 to 1.4 minutes. Also, the time for *mDiffExec* on 64 processors is fairly different from that on 16 and 32 processors. This appears to be caused by I/O load from other jobs running simultaneously with Montage. Additionally, since some of the modules' timings are increasing as the number of processors is increased, one would actually run the module on the number of processors that minimizes the timing. For example, *mBgExec* on this machine should only be run on 16 processors, no matter how many are used for the other modules.

These timings are probably close to the best that can be achieved on a single cluster, and can be thought of as a lower bound on any parallel implementation, including any grid implementation. However, there are numerous limitations to this implementation, including that a single pool of processors with shared access to a common file system is required, and that any single failure of a module or submodule will cause the entire job to fail, at least from that point forward. The Pegasus approach described in Section 6 can overcome these limitations.

### 7.6 Pegasus timing results

When using remote grid resources for the execution of the concrete workflow, there is a non-negligible overhead involved in acquiring resources and scheduling the computation over them. In order to reduce this overhead, Pegasus can aggregate the nodes in the concrete workflow into clusters so that the remote resources can be utilized more efficiently. The benefit of clustering is that the scheduling overhead (from Condor-G, DAGMan and remote schedulers) is incurred only once for each cluster. In the following results we cluster the nodes in the workflow within a workflow level (or workflow depth). In the case of Montage, the *mProject* jobs are within a single level, *mDiff* jobs are in another level, and so on. Clustering can be done dynamically based on the estimated run time of the jobs in the workflow and the processor availability.

Figure 8 shows the end-to-end time taken to create (*mDAG* and Pegasus) and execute (runtime) the concrete workflow to construct a 6 x 6 degree mosaic. As previously mentioned, Condor Glidein is used to acquire the resources. Once the resources are acquired, they are available for executing the workflow and there is no queuing delay at the remote resource. The workflow was executed using DAGMan running on a host at USC Information Sciences Institute. The time taken to transfer the input data and the output mosaic is not included in this figure. These measurements were made using Montage version 3.0β5. In this version *mDiff* and *mFitplane* are also available as a single module called *mDiffFit*, which has been used in the timing results shown.

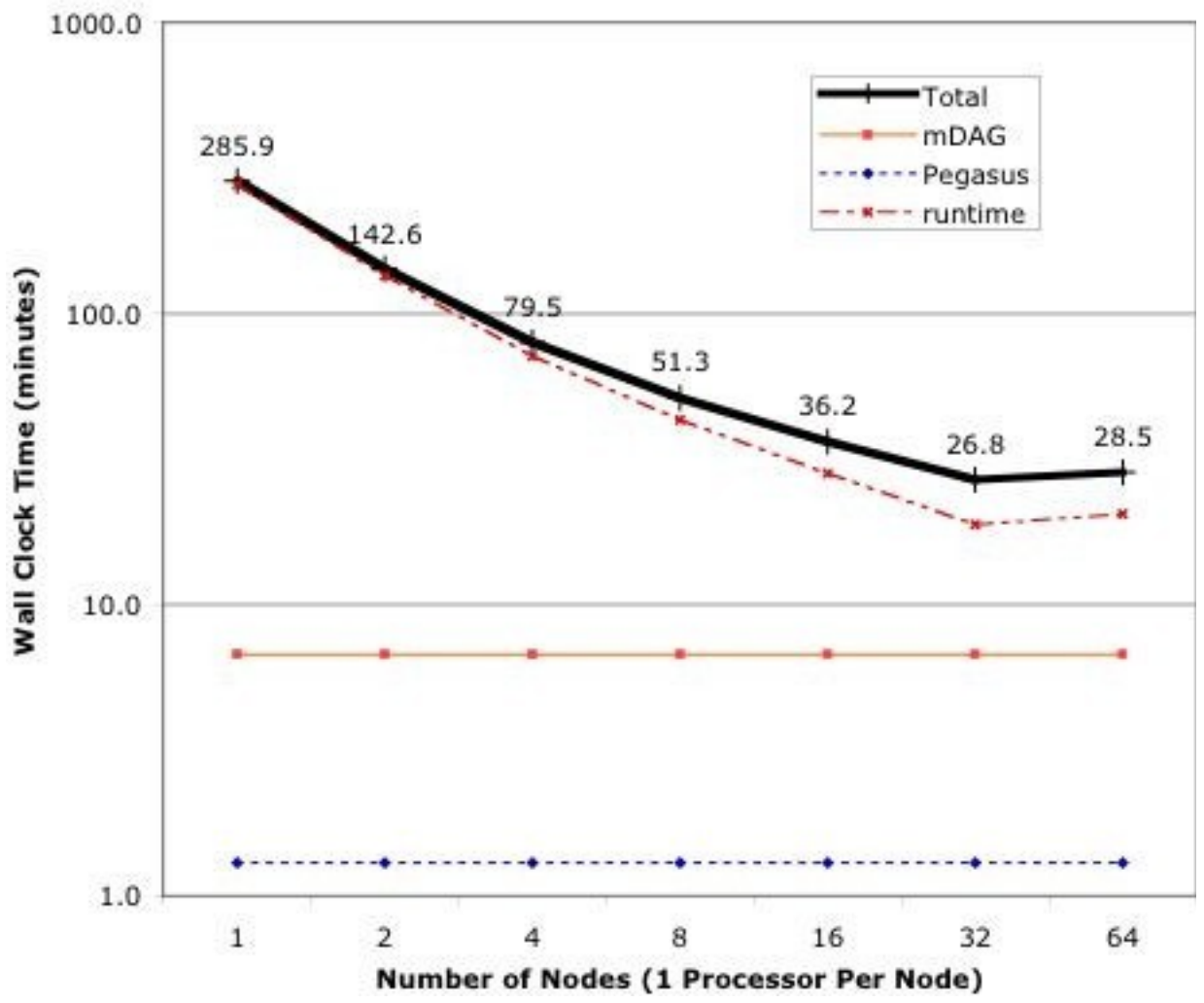

**Figure 8** *Times for building and executing the concrete workflow for creating a 6 x 6 degree mosaic.*

The figure shows the time in minutes for DAGMan to execute the workflow for different numbers of processors. The nodes in the workflow were clustered so that the number of clusters at each level of the workflow was equal to the number of processors. As the number of processors is increased (and thus the number of clusters increases), the Condor overhead becomes the dominant factor. DAGMan takes approximately 1 second to submit each cluster into the Condor queue. Condor's scheduling overhead adds additional delay. As a result we do not always see a corresponding decrease in the workflow execution time as we increase the number of processors. Also, as with the MPI results, the other codes running on the test machine appear to impact these timings. The 64-processor case seems to have worse performance than the 32-processor case, but it is likely that were it rerun on a dedicated machine, it would have better performance. This is discussed further below. Finally, there are sequential sections in the workflow that limit the overall parallel efficiency.

### 7.7 Timing discussion

Figure 9 shows a comparison of the time for the MPI run vs. the time needed to build and run the concrete DAG, for the benchmark problem. Notice that the performance of the Pegasus version seems to be faster than the MPI version except at 64 processors where the results are reversed. It is the authors' belief that, for large jobs, the measured difference between the Pegasus and MPI runs is not significant, and that it is due to the I/O contention caused by other jobs running on the test platform during these runs. A dedicated system would serve to mitigate these differences.

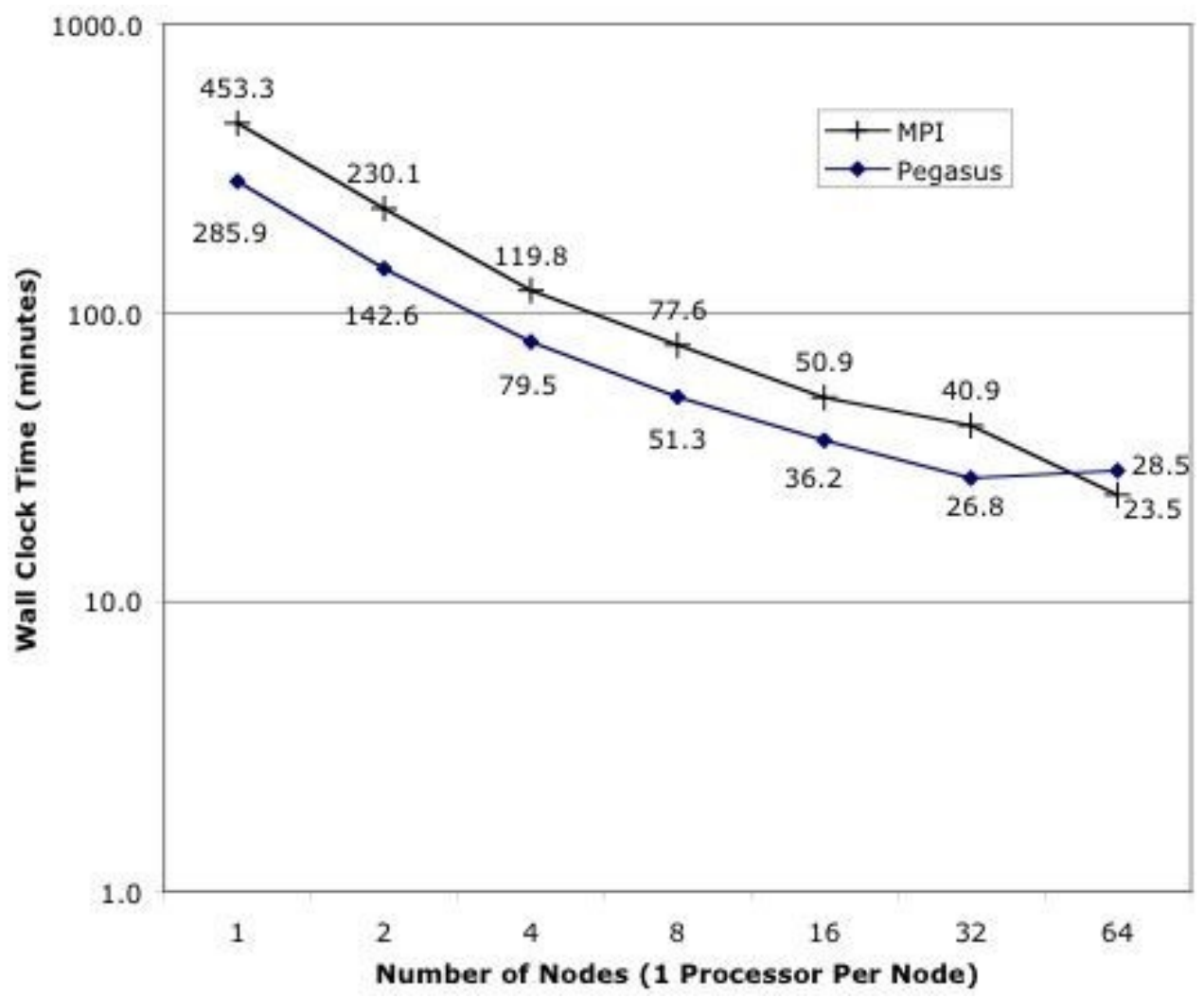

**Figure 9** *Times for building and executing the concrete workflow for creating a 6 x 6 degree mosaic.*

To examine some of these timings in more detail, we study the work needed to create a 1-degree square mosaic on 8 processors, as shown in Figure 10. The first difference is that *mImgtbl* is run three times in the MPI code vs. only once in the Pegasus code, where *mDAG* and Pegasus are run in advance instead of the first two *mImgtbl* runs. This is because the DAG is built in advance for Pegasus, but for MPI, the inputs are built on the fly from the files created by previous modules. Second, one MPI module starts immediately after the previous module finishes, while the Pegasus modules have a gap where nothing is running on the TeraGrid. This is the overhead imposed by DAGMan, as mentioned above. Third, the MPI code is almost 3 times faster for this small problem.

If we examine a larger problem, such as the 64 processor runs that create the 36 square degree test problem, as seen in Figure 11, we see some differences. First, the overall times are now comparable. Second, in the Pegasus case, the gaps between the modules are generally not noticeable, except between *mProject* and *mDiffFit* and between *mBgModel* and *mBackground*. Since the bars show the range of time of 64 processes now, some of the gaps are just hidden, while some are truly insignificant. Finally, in the Pegasus case, the *mDAG* and Pegasus times are substantial, but the *mAdd* time is much shorter than in the MPI case. Again, this is just a difference between the two implementations: *mDAG* allows the individual *mAdd* processes to open only the relevant files in the Pegasus case, whereas in the MPI case, the region of coverage is not known in advance, so all *mAdd* instances must open all files. Many are then closed immediately, if they are determined to not intersect the output tile. The I/O overhead in the MPI case is much larger, but the startup time is much shorter.

It is possible that a larger number of experiments run on a large dedicated machine would further illuminate the differences in performance between the MPI and Pegasus approaches, but even on the heavily-loaded TeraGrid cluster at NCSA, it is clear that there is no performance difference that outweighs the other advantages of the Pegasus approach, such as fault tolerance and the ability to use multiple machines for a single large job.

### 7.8 Finishing the job

Once the output mosaic has been built, it must be made available to the user, and the user must be notified of this availability. The Montage portal currently transfers the mosaic from the compute processors to the portal, and emails the user. In the case of Pegasus, the mosaic is also registered in RLS. The time required to transfer the mosaic and to notify the user are common to both the Pegasus and MPI approaches, and thus are not discussed here.

## 8 CONCLUSION

Montage was written as a very general set of modules to permit a user to generate astronomical image mosaics. A Montage mosaic is a single image that is built from multiple smaller images and preserves the photometric and astrometric accuracy of the input images. Montage includes modules that are used to reproject images to common coordinates, calculate overlaps between images, calculate coefficients to permit backgrounds of overlap regions to be matched, modify images based on those coefficients, and coadd images using a variety of methods of handling multiple pixels in the same output space.

The Montage modules can be run on a single processor computer using a simple shell script. Because this approach can take a long time for a large mosaic, alternatives to make use of the grid have been developed. The first alternative, using MPI versions of the computation-intensive modules, performs well but is somewhat limited. A second alternative, using Pegasus and other grid tools, is more general and allows for execution on a variety of platforms without requiring a change in the underlying code base, and appears to have real-world performance comparable to that of the MPI approach for reasonably large problems. Pegasus allows various scheduling techniques to be used to optimize the concrete workflow for a particular execution platform. Other benefits of Pegasus include opportunistically making best use of available resources through dynamic workflow mapping, and taking advantage of pre-existing intermediate data products.

The Montage software, user guide and user support system are available on the project web site at http://montage.ipac.caltech.edu/. Montage has been used by a number of NASA projects for science data product generation, quality assurance, mission planning, and outreach. These projects include: two Spitzer Legacy Projects, SWIRE (Spitzer Wide Area Infrared Experiment, http://swire.ipac.caltech.edu/swire/swire.html) and GLIMPSE (Galactic Legacy Infrared Mid-Plane Survey Extraordinaire, http://www.astro.wisc.edu/sirtf/); NASA's

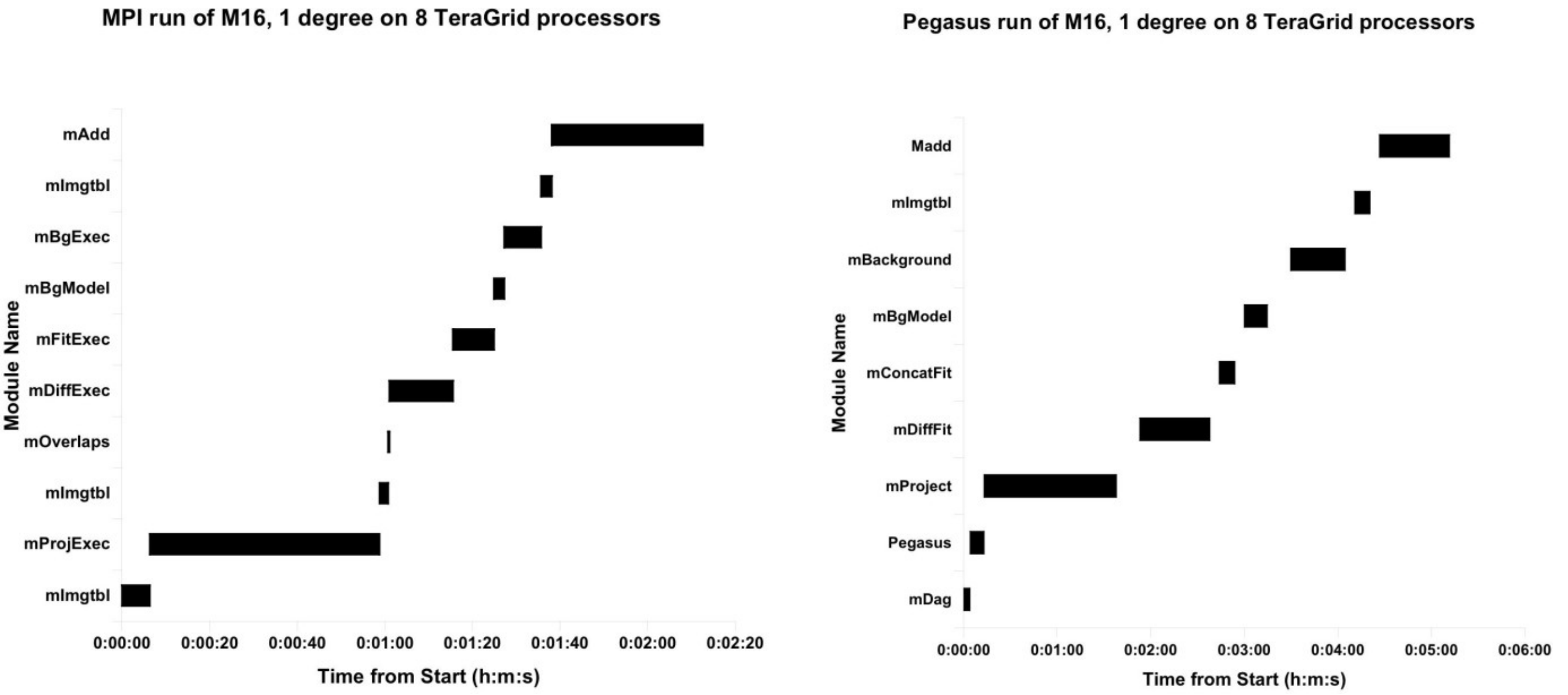

**Figure 10** *Timing of modules for creating a 1 x 1 degree mosaic on 8 processors. The MPI modules and timing are on the left, and the Pegasus modules and timing are on the right. The bars for the Pegasus modules show the start and end times of the first and last processes of each module.*

Infrared Science Archive, IRSA (http://irsa.ipac.caltech.edu/); and the Hubble Treasury Program, the COSMOS Cosmic Evolution Survey (http://www.astro.caltech.edu/~cosmos/).

## ACKNOWLEDGEMENT

Montage is funded by NASA's Earth Science Technology Office, Computational Technologies (ESTO-CT) Project, under Cooperative Agreement Number NCC5-626 between NASA and the California Institute of Technology. Pegasus is supported by NSF under grants ITR-0086044 (GriPhyN) and ITR AST0122449 (NVO).

Part of this research was carried out at the Jet Propulsion Laboratory, California Institute of Technology, under a contract with the National Aeronautics and Space Administration. Reference herein to any specific commercial product, process, or service by trade name, trademark, manufacturer, or otherwise, does not constitute or imply its endorsement by the United States Government or the Jet Propulsion Laboratory, California Institute of Technology.

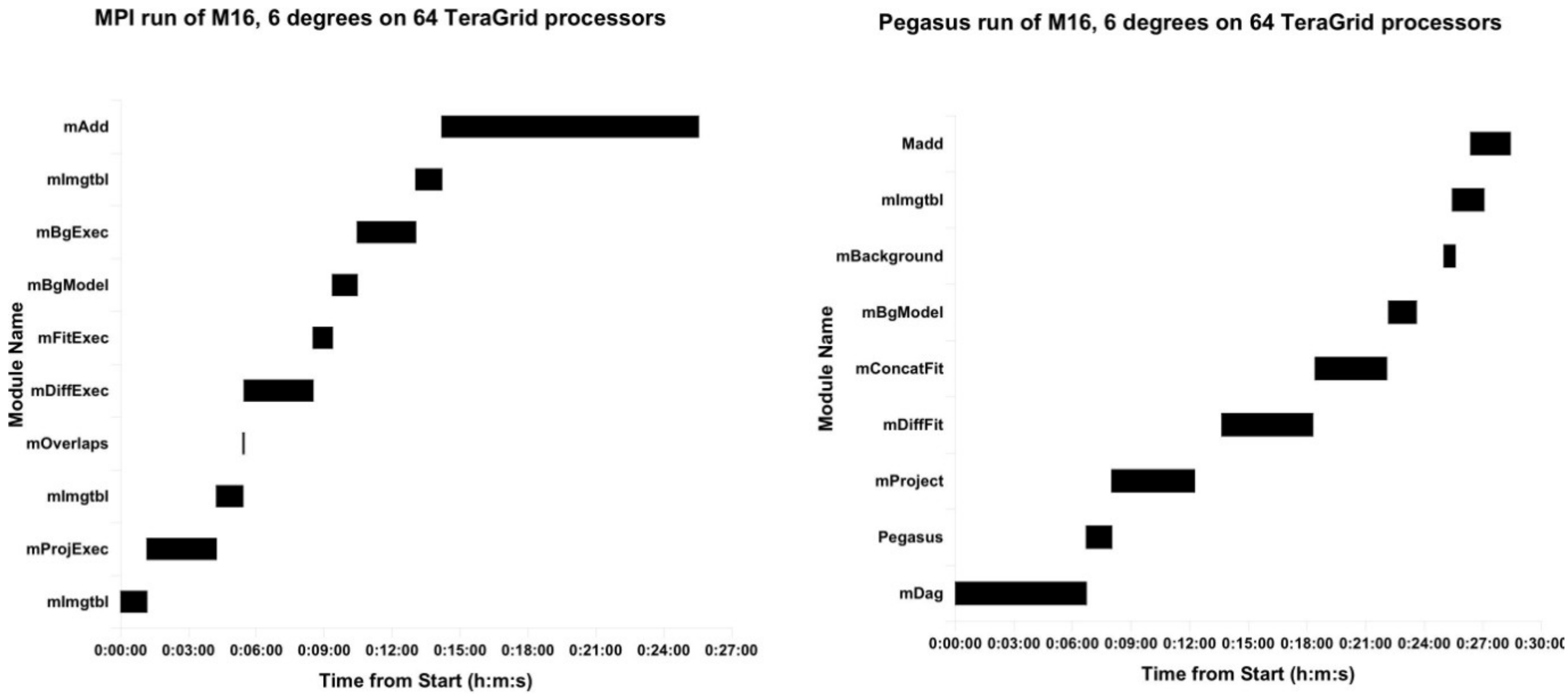

**Figure 11** *Timing of modules for creating an 8 x 8 degree mosaic on 64 processors. The MPI modules and timing are on the left, and the Pegasus modules and timing are on the right. The bars for the Pegasus modules show the start and end times of the first and last processes of each module.*

## WEBSITES